\documentclass[12pt]{article}
\usepackage{a4wide,epsfig}

\usepackage{slashed}

\newcommand{\agt}{\rlap{\lower 3.5 pt \hbox{$\mathchar \sim$}} \raise 1pt
 \hbox {$>$}}
\newcommand{\alt}{\rlap{\lower 3.5 pt \hbox{$\mathchar \sim$}} \raise 1pt
 \hbox {$<$}}
\newcommand{\re}{\mathop{\mathrm{Re}}\nolimits}

\catcode`@=11
\newcount\@tempcntc
\def\@citex[#1]#2{\if@filesw\immediate\write\@auxout{\string\citation{#2}}\fi
  \@tempcnta\z@\@tempcntb\m@ne\def\@citea{}\@cite{\@for\@citeb:=#2\do
    {\@ifundefined
       {b@\@citeb}{\@citeo\@tempcntb\m@ne\@citea\def\@citea{,}{\bf
?}\@warning
       {Citation `\@citeb' on page \thepage \space undefined}}%
    {\setbox\z@\hbox{\global\@tempcntc0\csname b@\@citeb\endcsname\relax}%
     \ifnum\@tempcntc=\z@ \@citeo\@tempcntb\m@ne
       \@citea\def\@citea{,}\hbox{\csname b@\@citeb\endcsname}%
     \else
      \advance\@tempcntb\@ne
      \ifnum\@tempcntb=\@tempcntc
      \else\advance\@tempcntb\m@ne\@citeo
      \@tempcnta\@tempcntc\@tempcntb\@tempcntc\fi\fi}}\@citeo}{#1}}
\def\@citeo{\ifnum\@tempcnta>\@tempcntb\else\@citea\def\@citea{,}%
  \ifnum\@tempcnta=\@tempcntb\the\@tempcnta\else
   {\advance\@tempcnta\@ne\ifnum\@tempcnta=\@tempcntb \else
\def\@citea{--}\fi
    \advance\@tempcnta\m@ne\the\@tempcnta\@citea\the\@tempcntb}\fi\fi}
\catcode`@=12
\begin{document}

\title{
\vskip-3cm{\baselineskip14pt
\centerline{\normalsize DESY 12--001\hfill ISSN 0418-9833}
\centerline{\normalsize December 2011\hfill}}
\vskip1.5cm
$J/\psi$ polarization at Tevatron and LHC: Nonrelativistic-QCD factorization
at the crossroads}

\author{Mathias Butenschoen, Bernd A. Kniehl\\
{\normalsize II. Institut f\"ur Theoretische Physik, Universit\"at Hamburg,}\\
{\normalsize Luruper Chaussee 149, 22761 Hamburg, Germany}
}
\date{}

\maketitle

\begin{abstract}
We study the polarization observables of $J/\psi$ hadroproduction at
next-to-leading order within the factorization formalism of nonrelativistic
quantum chromodynamics.
We complete the present knowledge of the relativistic corrections by also
providing the contribution due to the intermediate $^3\!P_J^{[8]}$ color-octet
states at this order, which turns out to be quite significant.
Exploiting the color-octet long-distance matrix elements previously extracted
through a global fit to experimental data of unpolarized $J/\psi$ production,
we provide theoretical predictions in the helicity and Collins-Soper frames
and compare them with data taken by CDF at Fermilab Tevatron I and II and by
ALICE at CERN LHC.
The notorious CDF $J/\psi$ polarization anomaly familiar from leading-order
analyses persists at the quantum level, while the situation looks promising
for the LHC, which is bound to bring final clarification. 

\medskip

\end{abstract}

\newpage
While the overly successful experiments at the LHC are closing in on the Higgs
boson and are systematically searching for signals of physics beyond the
standard model (SM), we must not be carried away losing track of a
longstanding, unresolved puzzle in quantum chromodynamics (QCD), the otherwise
well-established SU(3) gauge theory of the strong interactions, right in the
core of the SM.
In fact, despite concerted experimental and theoretical efforts ever since the
discovery of the $J/\psi$ meson in the November revolution of 1974 (The Nobel
Prize in Physics 1976), the genuine mechanism underlying the production and
decay of heavy quarkonia, which are QCD bound states of a heavy quark $Q=c,b$
and its antiparticle $\overline{Q}$, has remained mysterious.

Nonrelativistic QCD (NRQCD) \cite{Caswell:1985ui} endowed with an appropriate
factorization theorem, which was conjectured in a seminal work by Bodwin,
Braaten, and Lepage \cite{Bodwin:1994jh} and explicitly proven through
next-to-next-to-leading order \cite{Nayak:2005rt}, arguably constitutes the
most probable candidate theory at the present time.
This implies a separation of process-dependent short-distance coefficients, to
be calculated perturbatively as expansions in the strong-coupling constant
$\alpha_s$, from supposedly universal long-distance matrix elements
(LDMEs), to be extracted from experiment.
The relative importance of the latter can be estimated by means of velocity
scaling rules, which predict each of the LDMEs to scale with a definite
power of the heavy-quark velocity $v$ in the limit $v\ll1$.
In this way, the theoretical predictions are organized as double expansions in
$\alpha_s$ and $v$.
A crucial feature of this formalism is that the $Q\overline{Q}$ pair can at
short distances be produced in any Fock state
$n={}^{2S+1}L_J^{[a]}$ with definite spin $S$, orbital angular momentum
$L$, total angular momentum $J$, and color multiplicity $a=1,8$.
In particular, this formalism predicts the existence of intermediate
color-octet (CO) states in nature, which subsequently evolve into physical,
color-singlet (CS) quarkonia by the nonperturbative emission of soft gluons.
In the limit $v\to0$, the traditional CS model (CSM) is recovered in the case
of $S$-wave quarkonia.
In the case of $J/\psi$ production, the CSM prediction is based just on the
$^3\!S_1^{[1]}$ CS state, while the leading relativistic corrections, of
relative order ${\cal O}(v^4)$, are built up by the $^1\!S_0^{[8]}$,
$^3\!S_1^{[8]}$, and $^3\!P_J^{[8]}$ ($J=0,1,2$) CO states.

The CSM is not a complete theory, as may be understood by noticing that the NLO
treatment of $P$-wave quarkonia is plagued by uncanceled infrared
singularities, which are, however, properly removed in NRQCD.
In a way, NRQCD factorization \cite{Bodwin:1994jh}, appropriately improved at
large transverse momenta $p_T$ by systematic expansion in powers of
$m_Q^2/p_T^2$ \cite{Kang:2011zz}, is the only game in town, which makes its
experimental verification such a matter of paramount importance and general
interest \cite{Brambilla:2010cs}.

The experimental test of NRQCD factorization \cite{Bodwin:1994jh} has been
among the most urgent tasks on the agenda of the international quarkonium
community for the past fifteen years \cite{Brambilla:2010cs} and, with
high-quality data being so copiously harvested at the LHC, is now more
tantalizing than ever.
The present status of testing NRQCD factorization in charmonium production is
as follows.
As for the unpolarized $J/\psi$ yield, NRQCD factorization has recently been
consolidated at NLO by a global fit to the world's data of hadroproduction,
photoproduction, two-photon scattering, and $e^+e^-$ annihilation
\cite{Butenschoen:2011yh}, which successfully pinned down the three CO LDMEs in
compliance with the velocity scaling rules and impressively supported their
universality (see Table~1 in Ref.~\cite{Butenschoen:2011ks}).
On the other hand, the Tevatron II \cite{Acosta:2004yw} data alone can just fix
two linear combinations of the three CO LDMEs
\cite{Ma:2010yw,Butenschoen:2010px}, and the fit results of
Ref.~\cite{Ma:2010yw} are incompatible with Ref.~\cite{Butenschoen:2011yh}, as
discussed in Ref.~\cite{Butenschoen:2012qh}.
As for the $J/\psi$ polarization observables, a complete NLO NRQCD analysis has
so far only been performed for photoproduction \cite{Butenschoen:2011ks}.
The agreement with recent data from DESY HERA \cite{:2009br} was found to be
satisfactory at sufficiently large values of $p_T$, where nonperturbative
soft-gluon effects are negligible, provided that diffractive events were
eliminated by an appropriate acceptance cut.
The overall $\chi^2$ achieved at NLO in NRQCD turned out to be more than a
factor of 2 below the value obtained at NLO in the CSM
\cite{Artoisenet:2009xh}.
However, the case for NRQCD factorization appeared here to be not as
convincing as for the unpolarized $J/\psi$ yield \cite{Butenschoen:2011yh}.

In this Letter, we take the next---and possibly decisive---step in the
worldwide endeavor to put NRQCD factorization to the test and scrutinize
polarized $J/\psi$ hadroproduction at NLO in NRQCD.
Measurements by CDF in $p\overline{p}$ collisions with c.m.\ energy
$\sqrt{s}=1.8$~TeV \cite{Affolder:2000nn} and $\sqrt{s}=1.96$~TeV
\cite{Abulencia:2007us} at the Tevatron and by ALICE \cite{Abelev:2011md} in
$pp$ collisions with $\sqrt{s}=7$~TeV at the LHC are waiting to be rigorously
interpreted.
Previous NLO analyses were confined to the CSM \cite{Gong:2008sn} or included
only the $^1\!S_0^{[8]}$ and $^3\!S_1^{[8]}$ contributions \cite{Gong:2008ft},
which are deducible by well-established techniques.
We close this gap by providing also the $^3\!P_J^{[8]}$ contributions at NLO,
which are expected to be significant numerically.
Their calculation is far more intricate because the applications of the
respective projection operators to the short-distance scattering amplitudes
produce particularly lengthy expressions involving complicated tensor loop
integrals and exhibiting an entangled pattern of infrared singularities.
Technical details will be presented in a forthcoming publication.

The polarization of the $J/\psi$ meson is conveniently analyzed experimentally
by measuring the angular distribution of its leptonic decays, which is
customarily parametrized using the three polarization observables
$\lambda_\theta$, $\lambda_\phi$, and $\lambda_{\theta\phi}$, as
\begin{equation}
W(\theta,\phi)\propto1+\lambda_\theta\cos^2\theta
+\lambda_\phi\sin^2\theta\cos(2\phi)
+\lambda_{\theta\phi}\sin(2\theta)\cos\phi,
\end{equation}
where $\theta$ and $\phi$ are respectively the polar the azimuthal angles of
$l^+$ in the $J/\psi$ rest frame.
This definition depends on the choice of coordinate frame.
In the experimental analyses
\cite{Affolder:2000nn,Abulencia:2007us,Abelev:2011md}, the helicity and
Collins-Soper frames were employed, in which the polar axes point in the
directions of $-(\vec{p}_p+\vec{p}_{\overline{p}})$ and
$\vec{p}_p/|\vec{p}_p|-\vec{p}_{\overline{p}}/|\vec{p}_{\overline{p}}|$,
respectively.
The values $\lambda_\theta=0,+1,-1$ correspond to unpolarized, fully
transversely polarized, and fully longitudinally polarized $J/\psi$ mesons,
respectively.
In Refs.~\cite{Affolder:2000nn,Abulencia:2007us}, $\lambda_\theta$ is called
$\alpha$.
Working in the spin density matrix formalism and denoting the $z$ component of
$S$ by $i,j=0,\pm1$, we have
\begin{equation}
\lambda_\theta=\frac{d\sigma_{11}-d\sigma_{00}}{d\sigma_{11}+d\sigma_{00}},
\qquad
\lambda_\phi=\frac{d\sigma_{1,-1}}{d\sigma_{11}+d\sigma_{00}},
\qquad
\lambda_{\theta\phi}=
\frac{\sqrt{2}\re d\sigma_{10}}{d\sigma_{11}+d\sigma_{00}},
\label{eq:lam}
\end{equation}
where $d\sigma_{ij}$ is the $ij$ component of the $p\overline{p}\to J/\psi+X$
differential cross section.
An expression of $d\sigma_{ij}$ in terms of parton density functions (PDFs) and
partonic spin density matrix elements may be found in Eq.~(3) of
Ref.~\cite{Butenschoen:2011ks}.

In our numerical analysis, we adopt the CO LDME values from Table~I of
Ref.~\cite{Butenschoen:2011ks} along with
$\langle{\cal O}^{J/\psi}(^3\!S_1^{[1]})\rangle=1.32$~GeV$^3$
\cite{Bodwin:2007fz}, take the charm-quark mass, which we renormalize according
to the on-shell scheme, to be $m_c=1.5$~GeV, and use the one-loop (two-loop)
formula for $\alpha_s^{(n_f)}(\mu_r)$, with $n_f=4$ active quark flavors, at LO
(NLO).
As for the proton PDFs, we use the CTEQ6L1 (CTEQ6M) set \cite{Pumplin:2002vw}
at LO (NLO), which comes with an asymptotic scale parameter of
$\Lambda_\mathrm{QCD}^{(4)}=215$~MeV (326~MeV).
Our default choices for the $\overline{\mbox{MS}}$ renormalization,
factorization, and NRQCD scales are $\mu_r=\mu_f=m_T$ and $\mu_\Lambda=m_c$,
respectively, where $m_T=\sqrt{p_T^2+4m_c^2}$ is the $J/\psi$ transverse mass.
The theoretical uncertainty due to the lack of knowledge of corrections beyond
NLO is estimated by varying $\mu_r$, $\mu_f$, and $\mu_\Lambda$ by a factor 2
up and down relative to their default values.
In our NLO NRQCD predictions, we must also include the errors in the CO LDMEs,
which reflect the errors on the experimental data included in the fit.
To this end, we determine the maximum upward and downward shifts generated by
independently varying their values according to Table~I in
Ref.~\cite{Butenschoen:2011ks} and add the resulting half-errors in quadrature
to those due to scale variations.

\begin{figure*}
\begin{center}
\begin{tabular}{|c|c|c|}
\hline
\includegraphics[width=0.3\textwidth]{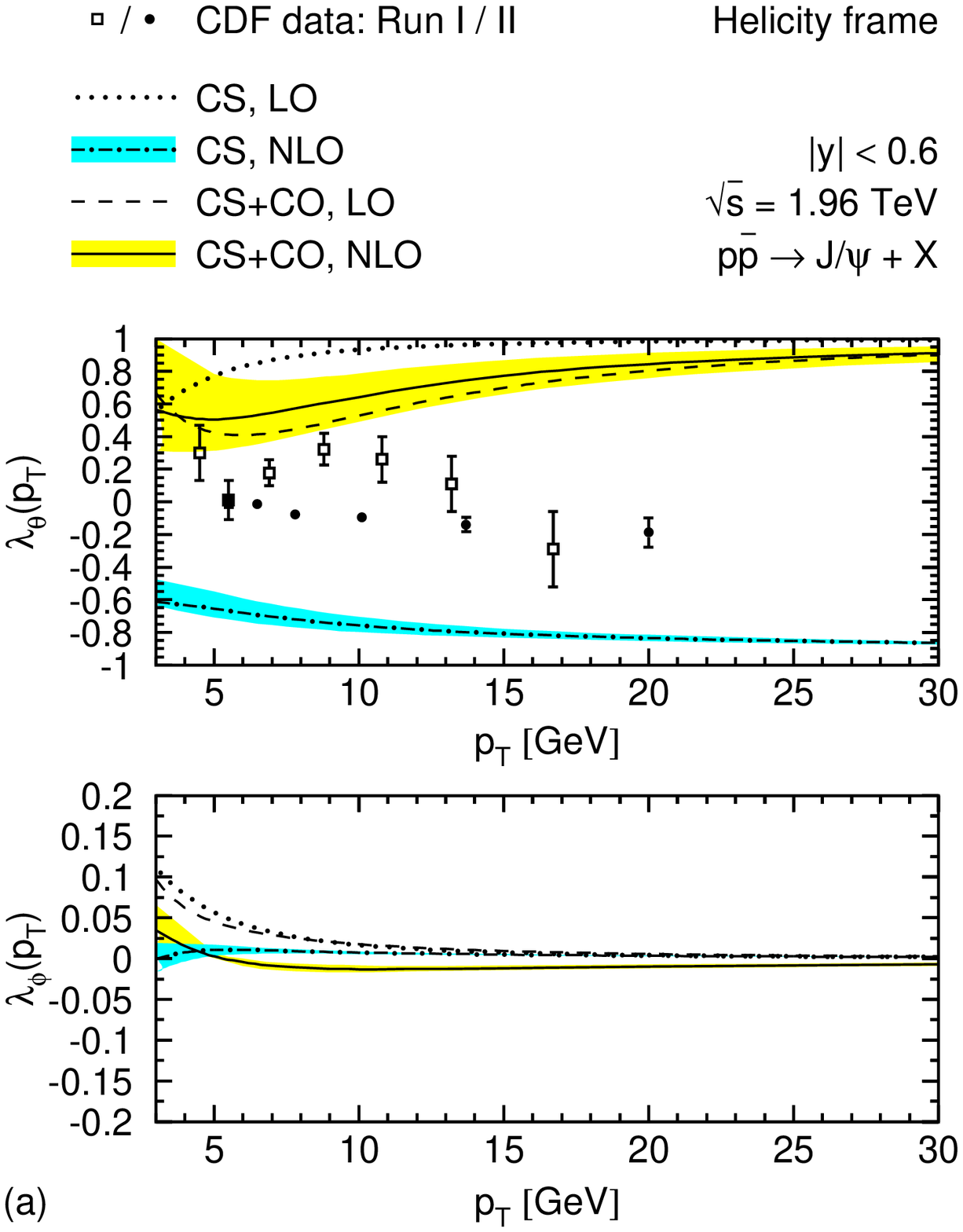}
&
\includegraphics[width=0.3\textwidth]{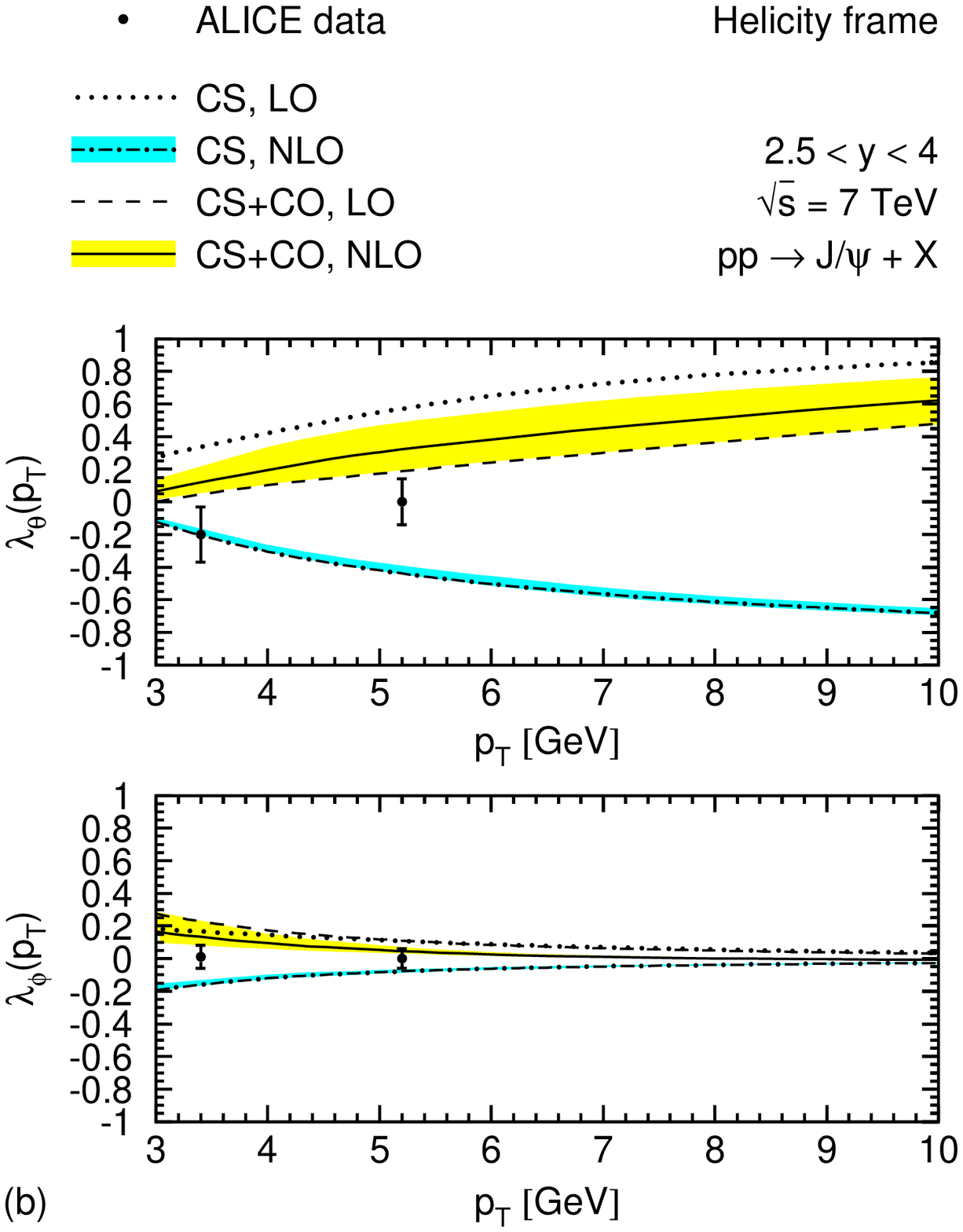}
&
\includegraphics[width=0.3\textwidth]{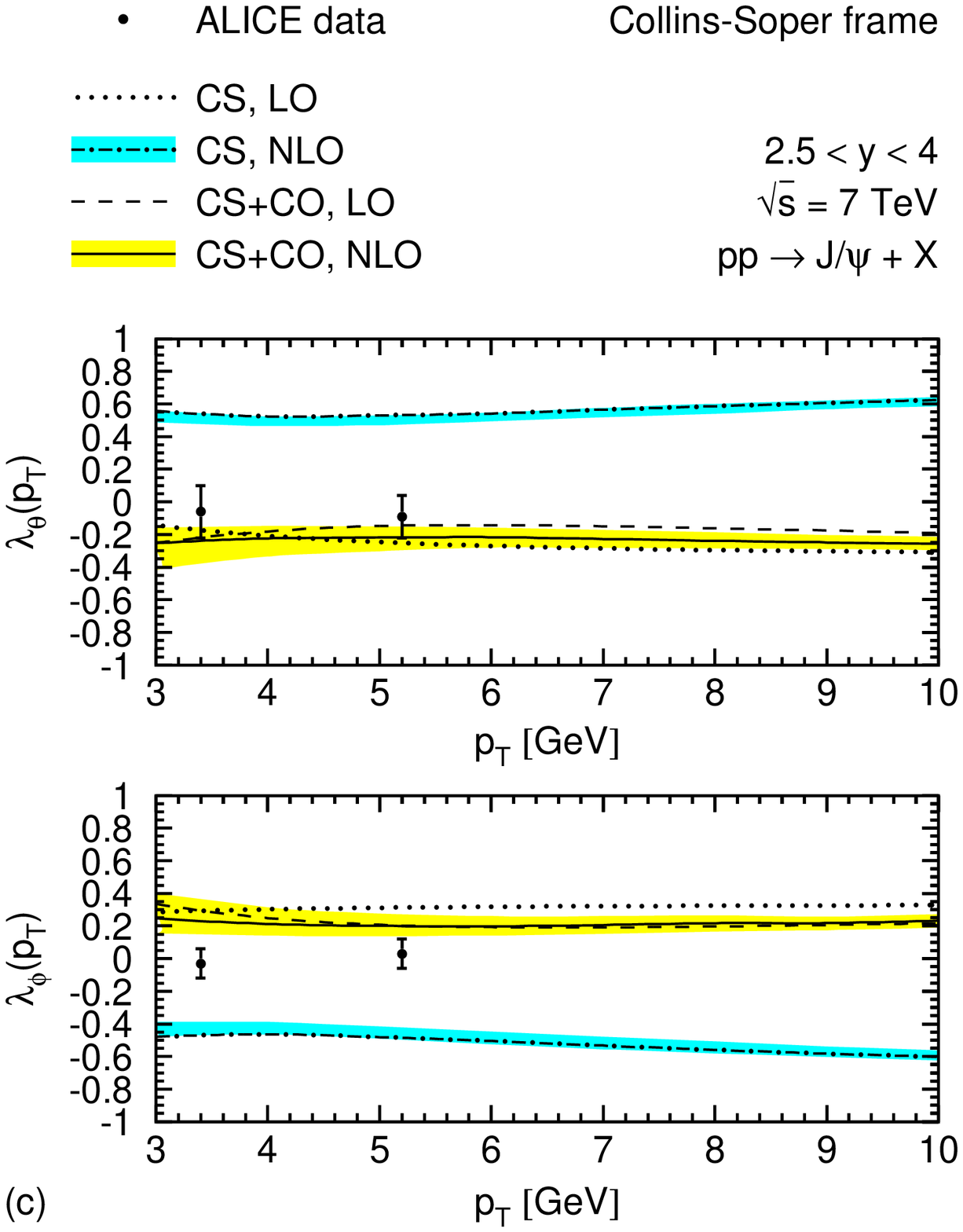}
\\
\hline
\end{tabular}
\end{center}
\caption{\label{fig:com}%
(color online) NLO NRQCD predictions (solid lines) for $\lambda_\theta$ and
$\lambda_\phi$ as functions of $p_T$ in the helicity and Collins-Soper frames
including theoretical uncertainties (shaded/yellow bands) compared to CDF
\cite{Affolder:2000nn,Abulencia:2007us} and ALICE \cite{Abelev:2011md} data.
For comparison, also the NLO CSM (dot-dashed lines) predictions including
theoretical uncertainties (hatched/blue bands) as well as the LO NRQCD (dashed
lines) and LO CSM (dotted lines) ones are shown.}
\end{figure*}

In Fig.~\ref{fig:com}, we confront our NLO NRQCD predictions for
$\lambda_\theta$ and $\lambda_\phi$ as functions of $p_T$ in the helicity and
Collins-Soper frames with the measurements by CDF 
\cite{Affolder:2000nn,Abulencia:2007us} and ALICE \cite{Abelev:2011md}.
Since the cross section ratios in Eq.~(\ref{eq:lam}) are very insensitive to
the precise value of $\sqrt{s}$, we may safely overlay the data from
$\sqrt{s}=1.8$~TeV \cite{Affolder:2000nn} with the predictions for
$\sqrt{s}=1.96$~TeV.
For comparison, also the LO NRQCD as well as the LO and NLO CSM predictions
are shown.
In order to visualize the size of the NLO corrections to the hard-scattering
cross sections, the LO predictions are evaluated with the same LDMEs.
As in Ref.~\cite{Butenschoen:2011yh}, we do not consider the range $p_T<3$~GeV,
where nonperturbative soft-gluon radiation invalidates a fixed-order treatment.
We observe that, in all the cases considered, the inclusion of the NLO
corrections has a considerably less dramatic effect in NRQCD than in the CSM,
where the normalizations and shapes of the various distributions are radically
modified.
This indicates that the perturbative expansion in $\alpha_s$ converges more
rapidly in NRQCD than in the CSM.
A similar observation has already been made for photoproduction
\cite{Butenschoen:2011ks}.
Comparing the LO NRQCD prediction for $\lambda_\theta$ in Fig.~\ref{fig:com}(a)
with the one referring to direct $J/\psi$ production in Fig.~1(b) of
Ref.~\cite{Braaten:1999qk}, we encounter marked differences in normalization
and shape, which mainly reflect the progress in our knowledge of the CO LDMEs
during the dozen years between Ref.~\cite{Braaten:1999qk} and
\cite{Butenschoen:2011yh}.
Comparing Figs.~\ref{fig:com}(a) and \ref{fig:com}(b), we observe that, for
$p_T\gg 2m_c$, the results are fairly stable when passing from the Tevatron to
the LHC and from the central ($|y|<0.6$) to the forward ($2.5<y<4$) rapidity
region, as long as one stays in the helicity frame.
However, switching from the helicity to the Collins-Soper frame has a radical
impact on the various $p_T$ distributions, as the comparison of
Figs.~\ref{fig:com}(b) and \ref{fig:com}(c) reveals.
The most striking effect appears for $\lambda_\theta$, where the NLO CSM and
NRQCD results are, roughly speaking, inverted.
As for $\lambda_\phi$, the Collins-Soper frame clearly outperforms the helicity
frame with regard to the power of NRQCD versus CSM discrimination.
As expected, the theoretical uncertainties due to scale variations steadily
decrease as the value of $p_T$ increases, which just reflects asymptotic
freedom.

Let us now compare experiment and theory.
The Tevatron~I data for $\lambda_\theta$ \cite{Affolder:2000nn} systematically
fall below the NLO NRQCD prediction and, unlike the latter, exhibit a downward
trend for $p_T\agt10$~GeV, but their errors are too large for a firm
conclusion.
This is quite different for the Tevatron~II data \cite{Abulencia:2007us}, which
are rather precise and indicate that the $J/\psi$ mesons are essentially
unpolarized in the helicity frame, while NRQCD clearly predicts transverse
polarization, with purity and precision steadily increasing with the value of
$p_T$.
A few caveats are in order here.
While the CDF data cover prompt production, including the feed-down from the
heavier $\chi_{cJ}$ and $\psi^\prime$ mesons, our predictions refer to direct
production, devoid of feed-down. 
However, the inclusion of feed-down was found to have a minor effect on the
$p_T$ distribution of $\lambda_\theta$ at LO in NRQCD \cite{Braaten:1999qk},
which is expected to carry over to NLO.
For $p_T<12$~GeV, the measurements from Tevatron runs I \cite{Affolder:2000nn}
and II \cite{Abulencia:2007us} are mutually incompatible, a feature that has
never been satisfactorily clarified by the CDF Collaboration.
Fortunately, the four LHC experiments are in a position to measure the
$J/\psi$ polarization with unprecedented precision, and ALICE has already
presented first results for $\lambda_\theta$ and $\lambda_\phi$
\cite{Abelev:2011md}, both in the helicity and Collins-Soper frames.
We compare them with our NLO NRQCD predictions in Figs.~\ref{fig:com}(b) and
\ref{fig:com}(c), respectively, leaving out one data point in the $p_T$ bin
2--3~GeV in each of the four cases.
The degree of agreement is encouraging.
All data points are at most one standard deviation away from the theoretical
error bands; four of them even overlap with the latter.
The agreement tends to improve with increasing value of $p_T$, as the
influence of uncontrolled corrections beyond NLO fades.
Except in the low-$p_T$ bin of $\lambda_\theta$ in the helicity frame, the
data clearly favor NRQCD over the CSM at NLO.
However, also these comparisons have to be taken with a grain of salt, since,
unlike our predictions, the ALICE data \cite{Abelev:2011md} also include
$J/\psi$ mesons from feed-down and $B$-meson decays.
The fraction of $J/\psi$ mesons from $b$-quark origin was measured by the LHCb
Collaboration for unpolarized production to be below 15\% in the $p_T$ range
considered here \cite{Aaij:2011jh}, so that their omission should have an
effect negligible against the theoretical uncertainty.

\begin{figure*}
\begin{center}
\begin{tabular}{ccc}
\includegraphics[width=0.3\textwidth]{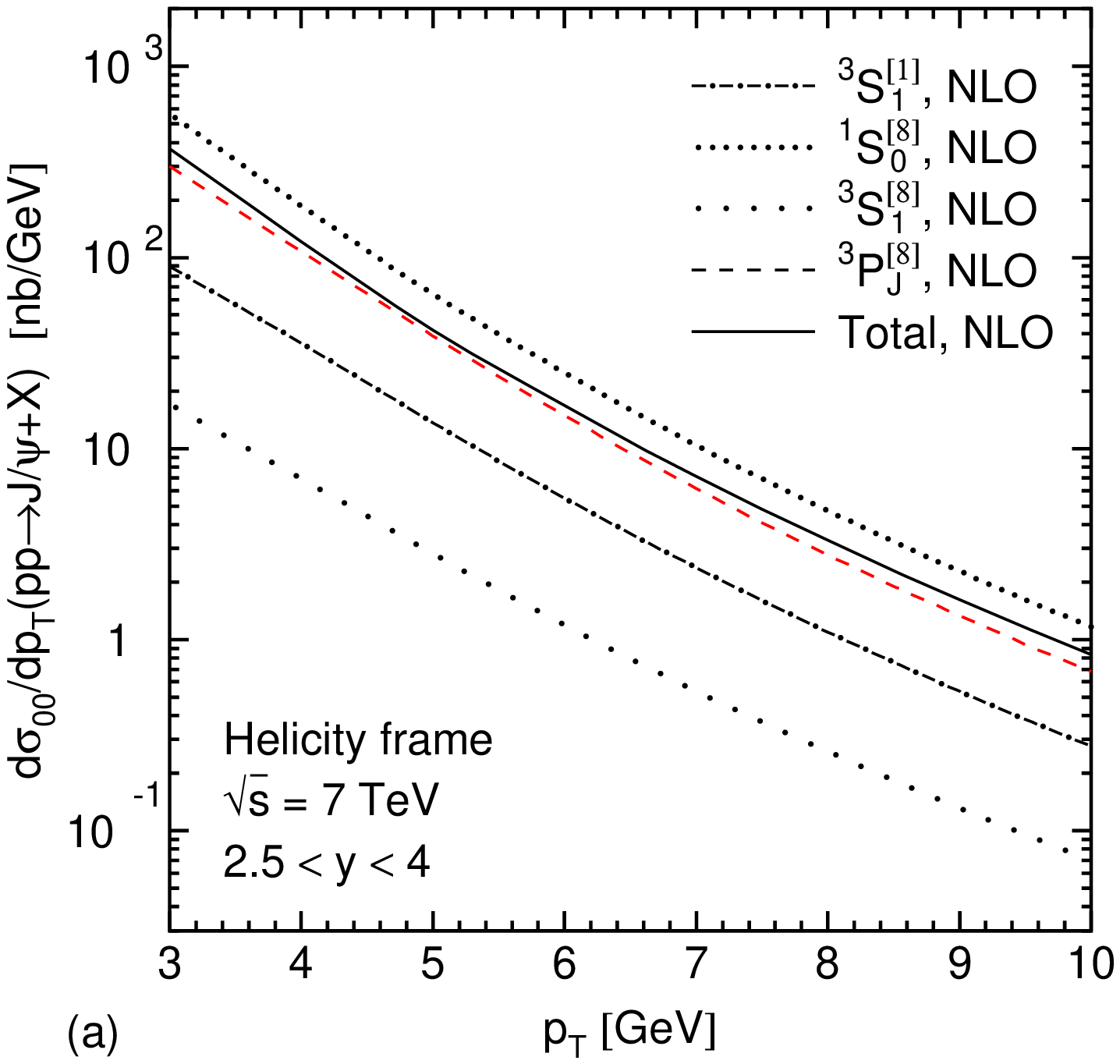}
&
\includegraphics[width=0.3\textwidth]{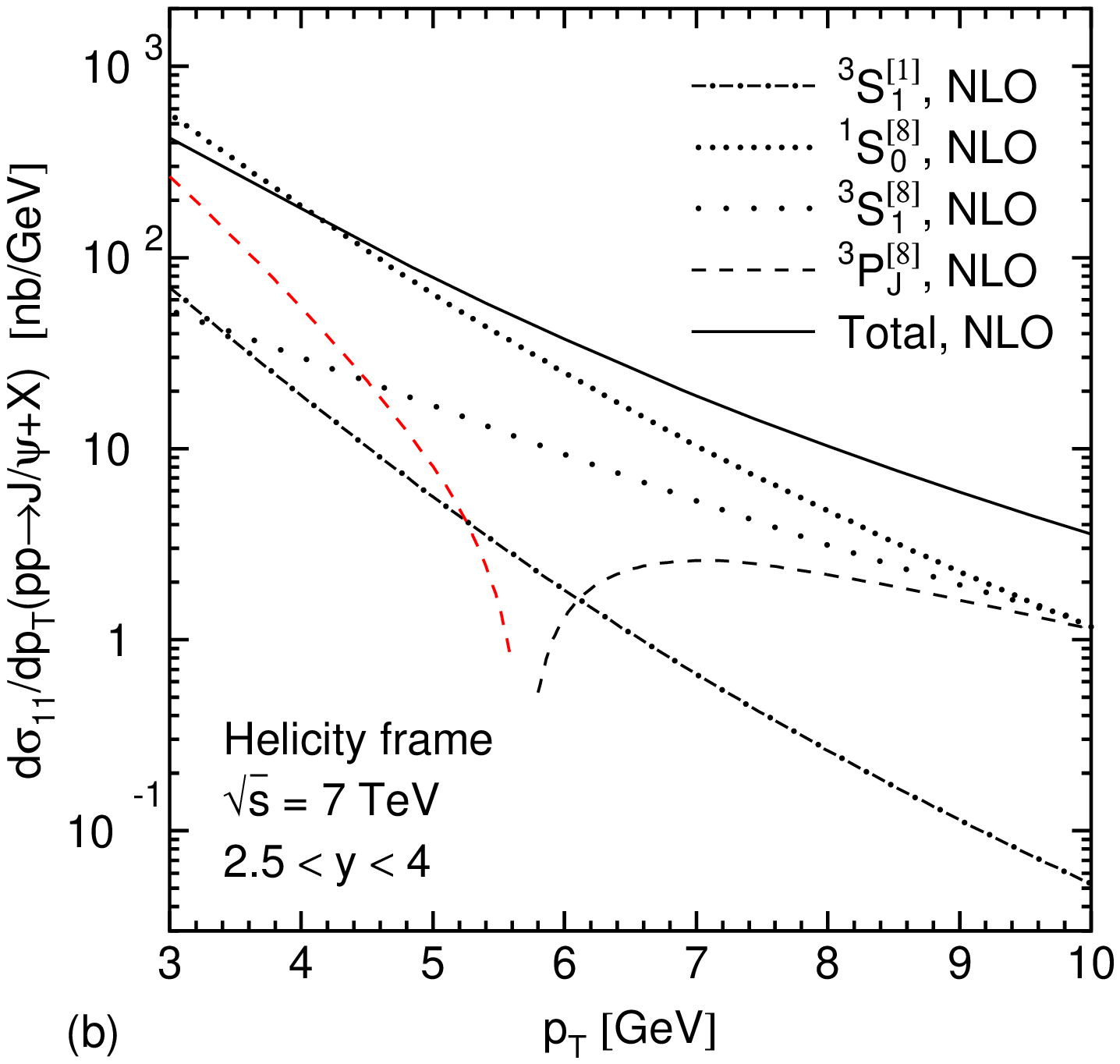}
&
\includegraphics[width=0.3\textwidth]{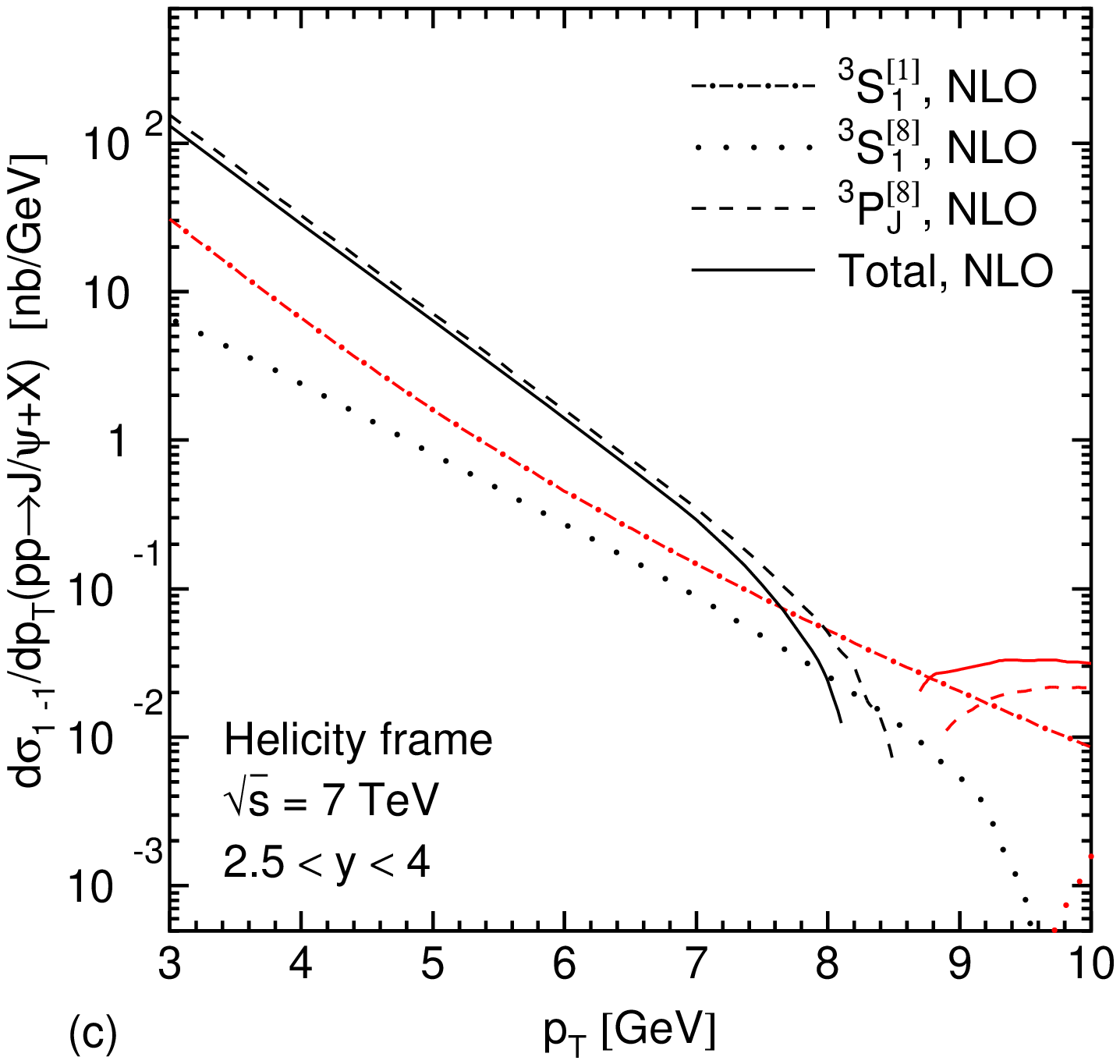}
\\
\includegraphics[width=0.3\textwidth]{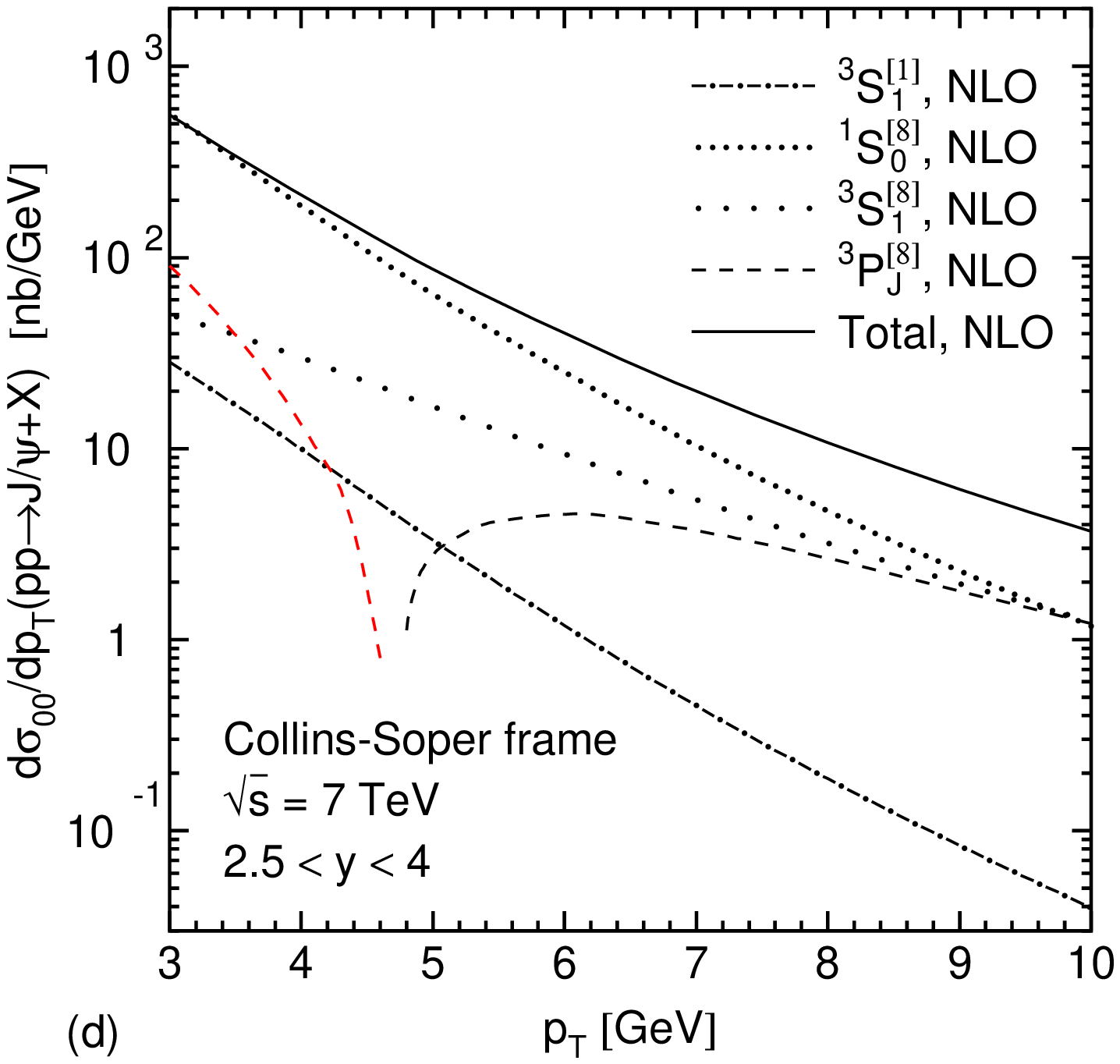}
&
\includegraphics[width=0.3\textwidth]{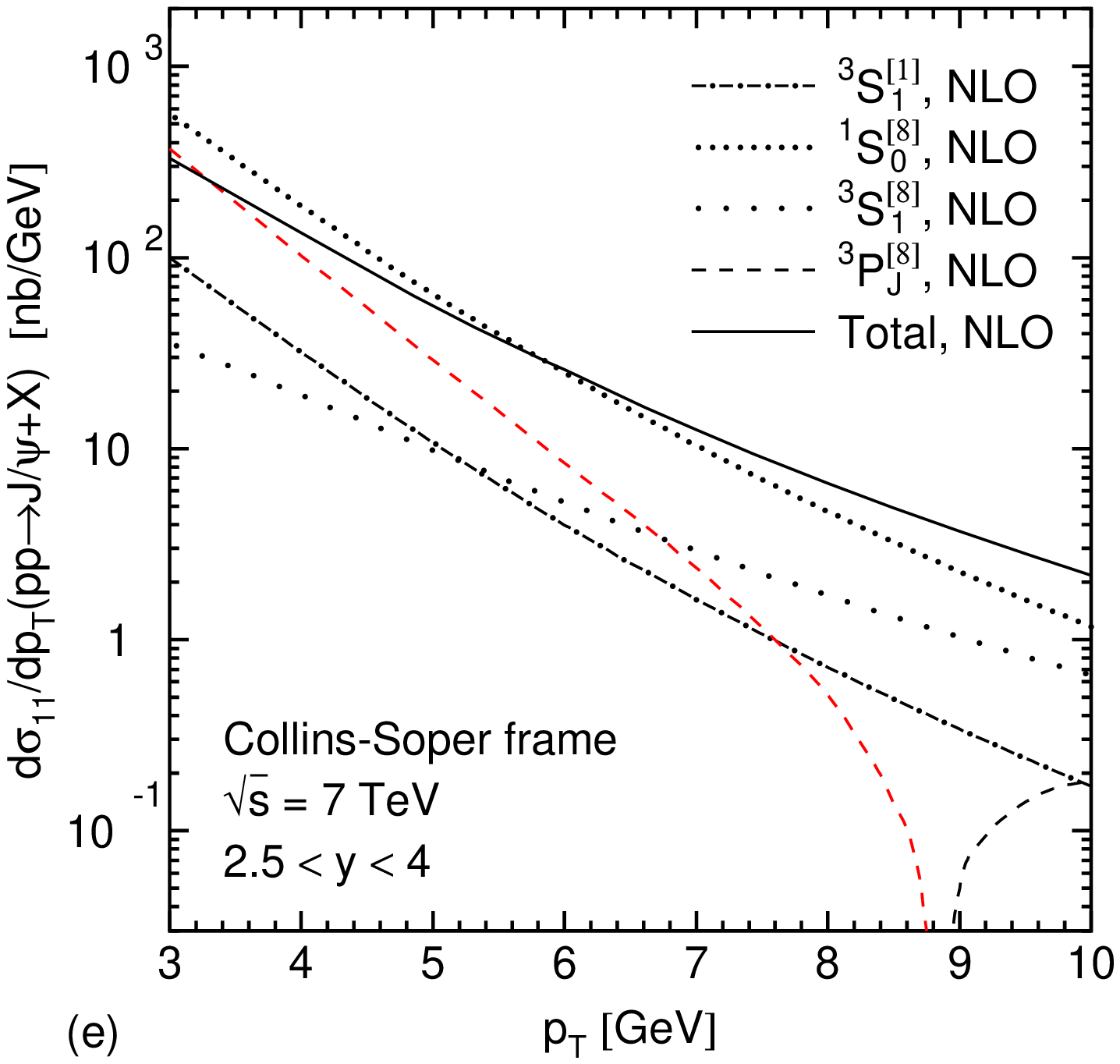}
&
\includegraphics[width=0.3\textwidth]{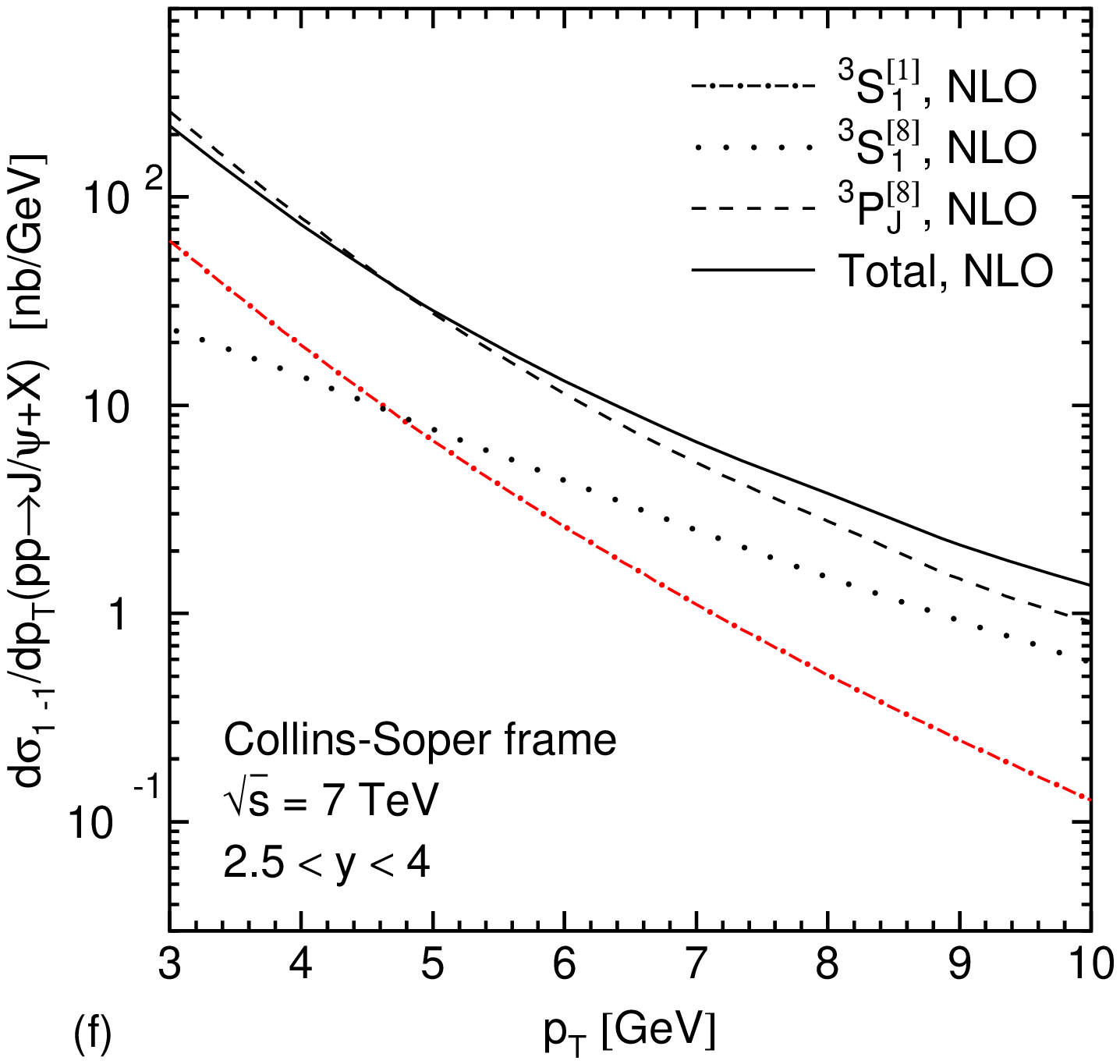}
\end{tabular}
\end{center}
\caption{\label{fig:den}%
(color online) $d\sigma_{00}/dp_T$, $d\sigma_{11}/dp_T$, and
$d\sigma_{1,-1}/dp_T$ for $pp\to J/\psi+X$ with $\sqrt{s}=7$~TeV and $2.5<y<4$
in the helicity and Collins-Soper frames at NLO in NRQCD (solid lines) and
their decompositions in the ${}^3\!S_1^{[1]}$ (dot-dashed lines),
${}^1\!S_0^{[8]}$ (dotted lines), ${}^3\!S_1^{[8]}$ (coarsely dotted lines),
and ${}^3\!P_J^{[8]}$ (dashed lines) contributions.
Negative values are marked red.}
\end{figure*}

In order to assess the relative importance of the individual $c\overline{c}$
Fock states $n$ for $\lambda_\theta$ and $\lambda_\phi$ in the helicity and
Collins-Soper frames, we detail their contributions to $d\sigma_{00}$,
$d\sigma_{11}$, and $d\sigma_{1,-1}$ in Fig.~\ref{fig:den}.
Note that the unpolarized cross section is recovered as
$d\sigma_{00}+2d\sigma_{11}$, while $d\sigma_{1,-1}$ is an auxiliary quantity,
which is entitled to take on either sign and receives no contribution from the
${}^1\!S_0^{[8]}$ channel.
As anticipated above, the previously unknown ${}^3P_J^{[8]}$ contributions
play a dominant role in this game, and their omission is bound to yield
meaningless results.
As is familiar from the unpolarized case \cite{Butenschoen:2011yh}, the
$^3\!P_J^{[8]}$ short-distance cross sections receive sizable NLO corrections
that may even turn them negative beyond some values of $p_T$, as may be
inferred from Figs.~\ref{fig:den}(b), (d), and (e) by recalling that our NLO
value of $\langle{\cal O}^{J/\psi}(^3\!P_0^{[8]})\rangle$
\cite{Butenschoen:2011yh} is also negative.
This is, however, unproblematic because a particular CO contribution
represents an unphysical quantity depending on the choices of renormalization
scheme and scale $\mu_\Lambda$ and is entitled to become negative as long as
the full cross section remains positive.

At this point, we compare our results with the theoretical literature.
We agree \footnote{%
In Eq.~(B58) of Ref.~\cite{Beneke:1998re}, the factor $(4\pi\alpha_s)$ should
be taken to the power 3.}
with the LO NRQCD results given in Eqs.~(B26)--(B62) of
Ref.~\cite{Beneke:1998re}.
We are able to nicely reproduce the LO and NLO ${}^3\!S_1^{[1]}$ and
${}^3\!S_1^{[8]}$ contributions to $\lambda_\theta$ shown as functions of
$p_T$ in Fig.~3 (9) of the first (second) paper in Ref.~\cite{Gong:2008sn}
and Fig.~4 of Ref.~\cite{Gong:2008ft}.

In conclusion, we presented the first complete NLO analysis of polarized
$J/\psi$ hadroproduction within NRQCD and, exploiting the knowledge of the CO
LDMEs from a global fit of unpolarized $J/\psi$ production data
\cite{Butenschoen:2011yh}, thus managed to establish certainty about that the
latest CDF measurement \cite{Abulencia:2007us} is in severe conflict with NRQCD
factorization \cite{Bodwin:1994jh}.
As emphasized in Ref.~\cite{Faccioli:2010kd}, the measurement of
$\lambda_\theta$ in the helicity frame alone provides only restricted access to
the rich phenomenology of $J/\psi$ polarization and might become ambiguous in
want of statistics, in particular if the patches of solid angle in which
simulations substitute data exceed a critical size.
In this sense, the ALICE Collaboration \cite{Abelev:2011md} has made a first
step into a bright future of heavy-quarkonium physics at the LHC, allowing for
ultimate tests of NRQCD factorization.

This work was supported in part by BMBF Grant No.\ 05H09GUE and HGF Grant No.\
HA~101.

{\it Note added.} After the submission of this Letter, a preprint
\cite{Chao:2012iv} appeared, in which $\lambda_\theta$ in the helicity frame
is studied at NLO in NRQCD for $J/\psi$ hadroproduction at the Tevatron.
Adopting the LDME values specified in the first line of Table~I in
Ref.~\cite{Chao:2012iv}, we are able to reproduce the numerical results of
Ref.~\cite{Chao:2012iv}.
Unfortunately, including the CDF data of
Ref.~\cite{Affolder:2000nn,Abulencia:2007us} on top of those from
Ref.~\cite{Acosta:2004yw}, already used in Ref.~\cite{Ma:2010yw}, still does
not allow for an independent determination of the three CO LDMEs, and the
result is violently incompatible with the global fit of
Ref.~\cite{Butenschoen:2011yh}.
In particular, this set of CO LDMEs, as well as the two alternative ones
specified in that table, are found \cite{Butenschoen:2012qh} to yield NLO-NRQCD
predictions for photoproduction which fail to describe the HERA data.
As a consequence, the conclusions of Ref.~\cite{Chao:2012iv} with regard to the
compliance of the CDF measurements of $\lambda_\theta$ in the helicity frame
\cite{Affolder:2000nn,Abulencia:2007us} with NRQCD factorization at NLO are
quite opposite to ours.

\end{document}